\documentclass[twocolumn,amsmath,amssymb]{revtex4}

\usepackage[dvips]{graphicx}
\usepackage[all]{xy}
\usepackage{amsmath}
\usepackage{amssymb}
\usepackage{dcolumn}
\usepackage{bm}
\usepackage{caption}

\begin{document}

\title{ CP Violating Phase Originated from Right-handed Neutrino Mixing}

\author{Xiao-Yan Wang}

\author{Xiang-Jun Chen}%
 \email{chenxj@hit.edu.cn}
\affiliation{%
Department of Physics, Harbin Institute of Technology,
Harbin, {\bf 150001},China 
}%

\date{\today}

\begin{abstract}
{\bf Abstract}--We  propose an idea that the observed CP violation in neutrino oscillation is originated from the  phase  in right-handed neutrino mixing by seesaw mechanism. We  add small breaking terms $M_{ij}N_{R_i}N_{R_j}  $  in model based on $A_4$ symmetry to  generate non-diagonal right-handed neutrino mixing, which will give a small correction to TBM mixing via seesaw mechanism with nonzero reactor angle and CP violating phase. We estimate the CP violating phase by investigating the process of leptogenesis due to the decay of right-handed neutrino.
\end{abstract}

\maketitle

Neutrino  oscillation \cite{sk,sno,chooz,dayabay} is one of the greatest discoveries in particle physics, which confirms neutrino mass and mixing. To explain the generation of neutrino mass, physicists consider neutrino as Majorana particle, which could generate mass term without right-handed neutrino which has not been sought up to now. Majorana mass term violates lepton number conservation which is required in the standard model(SM), thus, the discovery of neutrino oscillation is physics beyond SM.

Seesaw  mechanism is the most natural and elegant theory to explain the smallness of neutrino mass, which introduces very heavy  Majorana right-handed neutrino and leads to following relation
\begin{eqnarray}
M_v \approx - \frac{m_{\textrm{D}} m_{\textrm{D}}^T }{M_R },
\end{eqnarray}
where  $M_v $,  $m_{\textrm{D}}  $, $M_R $  are neutrino mass matrix, Dirac neutrino mass matrix and Majorana right-handed neutrino mass matrix, respectively. From Eq. (1), it is obvious that neutrino mass matrix $M_v $  will  be diagonalized  when $m_{\textrm{D}}  $ and  $M_R $  are both diagonalized.

Neutrino  mixing  is defined as
\begin{equation}
 \left(\begin{array}{c}
 v_e \\ v_{\mu} \\ v_{\tau} 
 \end{array} \right) = U  \left(\begin{array}{c}
 v_1 \\ v_2 \\ v_3 
 \end{array} \right),
 \end{equation}
where $v_l (l=e,\mu,\tau)  $  is flavor state which is defined by charged weak current, $v_i (i=1,2,3)  $ is neutrino mass eigenstate and $U$ is neutrino mixing matrix which is generally standard parameterized as PMNS matrix
\begin{eqnarray*}
 U_{\textrm{PMNS}} =\qquad \qquad \qquad   \qquad  \qquad  \qquad \qquad \qquad \qquad \qquad &&  \nonumber   \\
\left(\begin{array}{ccc}
 c_{12}c_{13} & s_{12}c_{13} & s_{13}  e^{-i\delta} \\
 -s_{12}c_{23}-c_{12}s_{13}s_{23}e^{i\delta} &  c_{12}c_{23}-s_{12}s_{13}s_{23}e^{i\delta} & c_{13}s_{23} \\
s_{12}s_{23}-c_{12}s_{13}c_{23}e^{i\delta} &  -c_{12}s_{23}-s_{12}s_{13}c_{23}e^{i\delta} & c_{13}c_{23}
 \end{array} \right)&&   \nonumber  \\
 \end{eqnarray*}
where $ s_{ij}=sin\theta_{ij},c_{ij}=cos\theta_{ij}  $  and $\delta=\delta_{\textrm{CP}}$  which is CP violating phase.  In 2002, P.F. Harrison et al. proposed TBM mixing pattern \cite{tbm} as follows
 \begin{equation}
 U_{\textrm{TBM}}=\left(\begin{array}{ccc}
 \frac{2}{\sqrt{6}} & \frac{1}{\sqrt{3}} & 0 \\
 -\frac{1}{\sqrt{6}} & \frac{1}{\sqrt{3}} & \frac{1}{\sqrt{2}} \\
 -\frac{1}{\sqrt{6}} & \frac{1}{\sqrt{3}} & -\frac{1}{\sqrt{2}}
 \end{array} \right),
 \end{equation}
which described neutrino oscillation very well. TBM mixing Eq. (3) indicates two large mixing angles  $ tan^2\theta_{12}=1/2, tan^2 \theta_{23}=1  $,   and zero reactor angle  $sin\theta_{13}=0  $. In 2012, Double Chooz \cite{chooz} and Daya Bay \cite{dayabay} experiments detected nonzero reactor angle  $ \theta_{13} $, which implied TBM mixing pattern should be modified to accord with experiments, while  TBM matrix is still considered as the best zeroth-order  neutrino mixing.  Authors exerted perturbation in various methods on TBM matrix to attempt to predict nonzero   $ \theta_{13} $. For example, literature
 \cite{2seesaw} generated deviation from TBM matrix with two seesaw mechanisms.

Moreover, nonzero reactor angle  $ \theta_{13} $ will give rise to the CP violation of neutrino oscillation.  Cabibbo \cite{cabibbo} and other physicists \cite{cpviolation} pointed out that three-flavor mixing would generate inevitable CP violating phase. For TBM mixing pattern with zero reactor angle, CP violating phase which accompanying  $ sin\theta_{13} $   would not act on neutrino oscillation. The detection of nonzero reactor angle urges experiments of detecting CP violating phase \cite{nova,t2k} and theories \cite{deviation} to predict it.

In this paper, we propose an idea that the observed nonzero reactor angle and CP violation is due to the entry with CP phase $ e^{i\delta} $ in  right-handed neutrino mixing via seesaw mechanism. Similar opinion was discussed in literature \cite{theory}. In our scheme, we will generate TBM mixing matrix in model based on discrete symmetry as the zeroth-order mixing.  According to Eq. (1), this TBM mixing could be considered as mixing in left-handed neutrino sector in Dirac mass term  with diagonal right-handed neutrino mixing matrix, which is only dependent of the way left-handed neutrino coupling to right-handed neutrino and  independent of right-handed neutrino mixing.  Then we will add softly breaking mass terms to generate non-diagonal right-handed neutrino mixing matrix, which will give a small correction to TBM matrix via seesaw mechanism. The resultant neutrino mixing will present nonzero reactor mixing angle and CP violating phase. At the end,  the CP violating phase will be estimated by investigating leptogenesis due to the decay of right-handed neutrino. 

The discrete flavor symmetry can control the structure of lepton mixing matrix. TBM mixing matrix is well known for its good fit for experiments, which is shown in Eq. (3)
 \begin{equation*}
 U_{\textrm{TBM}}=\left(\begin{array}{ccc}
 \frac{2}{\sqrt{6}} & \frac{1}{\sqrt{3}} & 0 \\
 -\frac{1}{\sqrt{6}} & \frac{1}{\sqrt{3}} & \frac{1}{\sqrt{2}} \\
 -\frac{1}{\sqrt{6}} & \frac{1}{\sqrt{3}} & -\frac{1}{\sqrt{2}}
 \end{array} \right).
 \end{equation*}
The neutrino mass matrix with TBM mixing can be decomposed into the sum of three simple matrices with integer elements which conform to certain discrete symmetry
\begin{eqnarray}
&&M_v=U^*_{\textrm{TBM}}\left(\begin{array}{ccc}
 m_1 & 0 & 0 \\
 0 & m_2 & 0 \\
0 & 0 & m_3
 \end{array} \right)U^+_{\textrm{TBM}}  \nonumber \\
&=&\frac{m_1+m_3   }{2}\left(\begin{array}{ccc}
 1 & 0 & 0 \\
 0 & 1 & 0 \\
0 & 0 & 1
 \end{array} \right) + \frac{m_2-m_1   }{2}\left(\begin{array}{ccc}
 1 & 1 & 1 \\
 1 & 1 & 1 \\
1 & 1 & 1
 \end{array} \right) \nonumber \\
&&+ \frac{m_1-m_3   }{2}\left(\begin{array}{ccc}
 1 & 0 & 0 \\
 0 & 0 & 1 \\
0 & 1 & 0
 \end{array} \right)
 \end{eqnarray}
where $m_1,m_2,m_3$  are neutrino masses.  The three matrices at the r. h. s of the second equality in Eq. (4) are all  $S_3$-symmetric. So any group including  $S_3$  subgroup and  triplet representation could generate TBM neutrino mixing matrix, in which the minimal groups are   $S_4$ and   $A_4$. 

$S_4$  group covers all permutations among four objects  $ (x_1,x_2,x_3,x_4)  $, i.e.  $ (x_1,x_2,x_3,x_4)\rightarrow (x_i,x_j,x_k,x_l)  $, which includes 24 elements and  non-Abelian subgroups:  $S_3$,   $A_4$ and $\Sigma(8)$ which is  $(Z_2 \times Z_2 ) \rtimes Z_2  $. The representations of $S_4$ include two singlets, one doublets and two triplets. In $S_4$  model, $S_4$  is spontaneously broken to one of subgroups, in which $Z_2 \times Z_2 $  is arranged for neutrinos which is called Klein four group and $Z_3$ (included in  $S_3$ and $A_4$ ) for charged leptons. The generators for these two subgroups, i.e.  $Z_2 \times Z_2 $ and $Z_3$  , could be chosen $d_1,a_2, b_1   $ (symbols used in appendix), which can be transformed by unitary matrix $U_w$  as
\begin{eqnarray}
&&X= U_w^+ d_1 U_w    \nonumber  \\
&=&\frac{1}{3}\left(\begin{array}{ccc}
 1 & 1 & 1 \\
 1 & w^* & w^{2*} \\
1 & w^{2*} & w^*
 \end{array} \right)\left(\begin{array}{ccc}
 1 & 0 & 0 \\
 0 & 0 & 1 \\
0 & 1 & 0
 \end{array} \right)\left(\begin{array}{ccc}
 1 & 1 & 1 \\
 1 & w & w^2 \\
1 & w^2 & w
 \end{array} \right)  \nonumber \\
&=&\left(\begin{array}{ccc}
 1 & 0 & 0 \\
 0 & 0 & 1 \\
0 & 1 & 0
 \end{array} \right),
 \end{eqnarray}

\begin{eqnarray}
S&=& U_w^+ a_2 U_w  \nonumber  \\
&=&\frac{1}{3}\left(\begin{array}{ccc}
 1 & 1 & 1 \\
 1 & w^* & w^{2*} \\
1 & w^{2*} & w^*
 \end{array} \right)\left(\begin{array}{ccc}
 1 & 0 & 0 \\
 0 & -1 & 0 \\
0 & 0 & -1
 \end{array} \right)\left(\begin{array}{ccc}
 1 & 1 & 1 \\
 1 & w & w^2 \\
1 & w^2 & w
 \end{array} \right)   \nonumber   \\
&=&\frac{1}{3}\left(\begin{array}{ccc}
 -1 & 2 & 2 \\
 2 & -1 & 2 \\
2 & 2 & -1
 \end{array} \right),
 \end{eqnarray}

\begin{eqnarray}
T&=& U_w^+ b_1 U_w   \nonumber   \\
&=&\frac{1}{3}\left(\begin{array}{ccc}
 1 & 1 & 1 \\
 1 & w^* & w^{2*} \\
1 & w^{2*} & w^*
 \end{array} \right)\left(\begin{array}{ccc}
 0 & 0 & 1 \\
 1 & 0 & 0 \\
0 & 1 & 0
 \end{array} \right)\left(\begin{array}{ccc}
 1 & 1 & 1 \\
 1 & w & w^2 \\
1 & w^2 & w
 \end{array} \right)    \nonumber    \\
&=&\left(\begin{array}{ccc}
 1 & 0 & 0 \\
 0 & w^2 & 0 \\
0 & 0 & w
 \end{array} \right),
 \end{eqnarray}
where $w^3=1$  and $U_w$  is called magic matrix
\begin{eqnarray}
 U_w=\frac{1}{\sqrt{3}}\left(\begin{array}{ccc}
 1 & 1 & 1 \\
 1 & w & w^2 \\
1 & w^2 & w
 \end{array} \right).
 \end{eqnarray}

The matrix which can diagonalize all generators Eq. (5), Eq. (6), Eq. (7) is only TBM matrix, which implies $S_4$  model could predict TBM mixing directly. Nevertheless, $S_4$  model is complicated and we will not choose it.

 $A_4$  group covers all even permutations of four objects, which is subgroup of $S_4$  and includes 12 elements. The  generators could be chosen $S=a_2, T=b_1$   which satisfy $ S^2=T^3=I $ and can be transformed by magic matrix  Eq. (8) as
\begin{eqnarray*}
S= U_w^+ a_2 U_w 
=\frac{1}{3}\left(\begin{array}{ccc}
 -1 & 2 & 2 \\
 2 & -1 & 2 \\
2 & 2 & -1
 \end{array} \right),
 \end{eqnarray*}

\begin{eqnarray*}
T= U_w^+ b_1 U_w   
=\left(\begin{array}{ccc}
 1 & 0 & 0 \\
 0 & w^2 & 0 \\
0 & 0 & w
 \end{array} \right).
 \end{eqnarray*}
Above two matrices  can both be diagonalized by TBM matrix, thus, TBM mixing matrix can be generated in $A_4$ model.   $A_4$ group includes three singlets and a single triplet, thus, we could put three left-handed neutrinos and charged leptons into the triplet, and put three right-handed charged leptons into three singlets as in literature \cite{ma}.
   More details are elaborated in appendix.

We adopt the $A_4$ model in literature \cite{ma} to generate the zeroth-order TBM neutrino mixing. The model introduced very heavy right-handed neutrinos and the assignment of leptons was
\begin{eqnarray}
&&(v_i, l_i )_L \ \  (\underline{3},1) \qquad N_{iR} \  \    (\underline{3},1)   \nonumber \\
&&l_{1R}  \ \    (\underline{1},1)\qquad \qquad l_{2R} \ \ (\underline{1'},1)  \qquad  \qquad l_{3R}   \ \  (\underline{1''},1)  \nonumber  \\
&&\Phi_i = (\phi_i^+, \phi_i^0 ) \ \   (\underline{3},0) \qquad 
\eta = (\eta^+, \eta^0 ) \ \ (\underline{1},-1).
 \end{eqnarray}
Lepton  masses are generated by the following Lagrangian
\begin{eqnarray}
\mathcal{L}& = &\frac{1}{2} M N^2_{iR} + f \overline{N}_{iR} (v_{iL}  \eta^0  -l_{iL} \eta^+ )  \nonumber   \\
&& + h_{ijk} \overline{(v_i,l_i)_L } l_{jR} \Phi_k + h.c.
 \end{eqnarray}
The model will generate TBM neutrino mixing  by seesaw mechanism. In  literature \cite{ma}, the author  added softly breaking terms $ m_{ij}N_{R_i}N_{R_j}(i\ne j) $   to remove neutrino mass degeneration, then the diagonal neutrino mass matrix could be expressed as
\begin{eqnarray}
Diag M^0_v\approx - \frac{ U_{\textrm{TBM}}^T m_{\textrm{D}} m_{\textrm{D}}^T U_{\textrm{TBM}}   }{M^0_R  }, 
\end{eqnarray}
where right-handed neutrino mass matrix $M_R^0$  could be considered diagonal. As a real unitary matrix  $U_{TBM}^T=U_{TBM}^+  $, TBM matrix $ U_{TBM} $  could be considered as mixing  in left-handed neutrino sector in Dirac mass term, which was only dependent of the way left-handed neutrino coupling to right-handed neutrino and  independent of right-handed neutrino mixing.

Then we  further  add very small breaking terms $ M_{ij}N_{R_i}N_{R_j}(i\ne j) $    as perturbation to generate non-diagonal right-handed neutrino mixing, in which we assume   $sin\theta_{12}(N_R)\sim sin\theta_{23}(N_R)\sim sin\theta_{13}(N_R)=sin\alpha \sim 0   $.    Now  Eq. (11) becomes non-diagonal again
\begin{eqnarray}
 - \frac{ U_{\textrm{TBM}}^+ m_{\textrm{D}} m_{\textrm{D}}^T U_{\textrm{TBM}}   }{ M_R   }
\end{eqnarray}
which will be diagonalized by $U_R$  as 
\begin{eqnarray}
Diag M_v & =& U^T M_v U \approx - \frac{ U_{\textrm{TBM}}^+ m_{\textrm{D}} m_{\textrm{D}}^T U_{\textrm{TBM}}   }{U_R^T M_R U_R   } \nonumber \\
&= &- U_R^*\frac{ U_{\textrm{TBM}}^+ m_{\textrm{D}} m_{\textrm{D}}^T U_{\textrm{TBM}}   }{ M_R   }U_R^+,
\end{eqnarray}
where
\begin{eqnarray}
M_v\approx - \frac{  m_{\textrm{D}} m_{\textrm{D}}^T   }{ M_R   }.
\end{eqnarray} 
Then neutrino mixing matrix will be   $U= U_{\textrm{TBM}}U_R^+   $, where we have used   $U_{\textrm{TBM}}^T=U_{\textrm{TBM}}^+  $.

For  right-handed neutrino mixing is approximately diagonal, we use the following simple matrix as approximation
\begin{equation}
 U_{\textrm{R}}\sim \left(\begin{array}{ccc}
 cos \alpha & 0 & sin \alpha e^{-i\delta} \\
 0 & 1 & 0 \\
-sin \alpha e^{i\delta} & 0 & cos \alpha
 \end{array} \right).
 \end{equation}
The  phase in Eq. (14) can not be absorbed in $N_R$  due to its Majorana property, i.e.  $N_R\equiv N_R^c$. Then neutrino mixing matrix can be written approximately as
 \begin{eqnarray}
 &&U=U_{\textrm{TBM}} U_R^+     \nonumber   \\
&=&\left(\begin{array}{ccc}
 \frac{2}{\sqrt{6}} & \frac{1}{\sqrt{3}} & 0 \\
 -\frac{1}{\sqrt{6}} & \frac{1}{\sqrt{3}} & \frac{1}{\sqrt{2}} \\
 -\frac{1}{\sqrt{6}} & \frac{1}{\sqrt{3}} & -\frac{1}{\sqrt{2}}
 \end{array} \right)\left(\begin{array}{ccc}
 cos \alpha & 0 & -sin \alpha e^{-i\delta} \\
 0 & 1 & 0 \\
sin \alpha e^{i\delta} & 0 & cos \alpha
 \end{array} \right)  \nonumber  \\
&=& \left(\begin{array}{ccc}
 \frac{2}{\sqrt{6}}cos\alpha & \frac{1}{\sqrt{3}} & -\sqrt{\frac{2}{3}}sin \alpha e^{-i\delta} \\
 -\frac{1}{\sqrt{6}}cos \alpha + \frac{1}{\sqrt{2}}sin\alpha e^{i\delta} & \frac{1}{\sqrt{3}} & \frac{1}{\sqrt{6}}sin \alpha e^{-i\delta}+\frac{1}{\sqrt{2}}cos \alpha  \\
 -\frac{1}{\sqrt{6}}cos \alpha - \frac{1}{\sqrt{2}}sin \alpha e^{i\delta} & \frac{1}{\sqrt{3}} & \frac{1}{\sqrt{6}}sin \alpha e^{-i\delta}-\frac{1}{\sqrt{2}}cos \alpha
 \end{array} \right).   \nonumber \\
 \end{eqnarray}  
In Eq. (15), TBM mixing matrix is slightly modified with nonzero reactor angle and CP violating phase, from which we can quickly obtain relation about parameter  $ \alpha $
\begin{equation}
sin\theta_{13}=-\sqrt{ \frac{2}{3}   }sin\alpha.
 \end{equation}
We  notice that in our scheme, the CP violating phase in neutrino mixing is the same as the one in right-handed neutrino mixing.

In 1986, Fukugita et al. \cite{fukugita} proposed that the origin of cosmological baryon number asymmetry or lepton number asymmetry was leptogenesis due to the CP violation in the decay of Majorana right-handed neutrino.  By analogy with the opinion in literature \cite{theory}, we assume that CP violation in decay of Majorana right-handed neutrino is due to CP violating phase in right-handed neutrino mixing.

According to Eq. (10), the Lagrangian of leptogenesis due to the decay of right-handed neutrino is
\begin{equation}
 \mathcal{L}_{leptogenesis} = f_{ij} \overline{N}_{iR} l_{jL} \eta^+  + h.c.
 \end{equation}
Right-handed neutrino Majorana mass term is
\begin{equation}
 \left(\begin{array}{ccc}
N_{eR}^T & N_{\mu R}^T & N_{\tau R}^T
 \end{array} \right) M_R  \left(\begin{array}{c}
 N_{eR} \\N_{\mu R} \\ N_{\tau R}
 \end{array} \right),
 \end{equation}
where  mass matrix $M_R  $ can be diagonalized as
\begin{equation}
 \left(\begin{array}{ccc}
N_{eR}^T & N_{\mu R}^T & N_{\tau R}^T
 \end{array} \right)U^*_R Diag M_R U^+_R \left(\begin{array}{c}
 N_{eR} \\N_{\mu R} \\ N_{\tau R}
 \end{array} \right).
 \end{equation}
   Then the decay of $N_R  $  will be
\begin{equation}
 \left(\begin{array}{ccc}
\overline N_{eR} &\overline N_{\mu R} &\overline N_{\tau R}
 \end{array} \right)U_R\eta^+ \left(\begin{array}{c}
 l_{eL} \\l_{\mu L} \\ l_{\tau L}
 \end{array} \right).
 \end{equation}
The Lagrangian of leptogenesis  Eq. (17) leads to 
\begin{equation}
f_{ij} \overline{N}_{iR} l_{jL} \eta^+= \left(\begin{array}{ccc}
\overline N_{eR} &\overline N_{\mu R} &\overline N_{\tau R}
 \end{array} \right)F\eta^+ \left(\begin{array}{c}
 l_{eL} \\l_{\mu L} \\ l_{\tau L}
 \end{array} \right),
 \end{equation}
where  $F$  is the coupling matrix as
\begin{equation}
F=\left(\begin{array}{ccc}
 f_{11} & f_{12} & f_{13} \\
 f_{21} &f_{22} & f_{23} \\
f_{31}& f_{32} & f_{33}
 \end{array} \right) 
 \end{equation}
which indicates the couplings to $N_{eR}$  are  $ f_{11} , f_{12} , f_{13}  $, and the rest can be done in a similar fashion. Compared Eq. (21) with Eq. (20) , coupling matrix  $F$  could be rewritten as coupling strength multiplying  $U_R$ 
\begin{equation}
F=\left(\begin{array}{ccc}
 f_{11} & f_{12} & f_{13} \\
 f_{21} &f_{22} & f_{23} \\
f_{31}& f_{32} & f_{33}
 \end{array} \right) =\left(\begin{array}{ccc}
 f_{1} &  &  \\
 &f_{2} &\\
&  & f_{3}
 \end{array} \right)U_R,
 \end{equation}
where  $f_1,f_2, f_3  $  are coupling strengths. We assume  $M_1 < M_2 < M_3  $. The  strength of a fermion coupling to certain scalar   field is proportional to mass of the fermion and then we have  $f_1 < f_2 < f_3  $. According to literature\cite{fukugita} , we choose $f_1 \sim 10^{-5}  $   and    $M_1 \sim 10^4 \textrm{GeV}$, then  $1\ge f_2,f_3 > 10^{-5}$.

The net lepton number produced by the decay of right-handed neutrino can be calculated by
\begin{equation}
\epsilon = \frac{9}{4\pi}\textrm{Im}(f_{ij}f_{jk}f_{il}^*f_{lk}^*    )I(M^2_l / M^2_k)/(FF^+)_{11},
\end{equation}
where
\begin{equation}
I(x)=x^{1/2}\{1+(1+x)ln[x/(1+x)]\}.
\end{equation}
Eq. (24) can lead to
\begin{eqnarray}
\epsilon &=& \frac{9}{4\pi}\textrm{Im}(f_{12}f_{23}f_{13}^*f_{33}^*    )I(M^2_3 / M^2_3)/(FF^+)_{11}  \nonumber   \\
\rightarrow \epsilon &=& \frac{9}{4\pi}f_1^2 f_2 f_3 \textrm{Im}((U_R)_{12}(U_R)_{23}(U_R)_{13}^*(U_R)_{33}^*    ) \nonumber  \\
&&\times I(1)/(FF^+)_{11},
\end{eqnarray}
where
\begin{equation}
I(M^2_3 / M^2_3)=I(1)=-0.3863.
\end{equation}
Taking $\epsilon < 10^{-6} $, $ (FF^+)_{11}\sim 10^{-10} $\cite{fukugita}, Eq. (26) will be
\begin{eqnarray}
&& -f_2 f_3 \textrm{Im}((U_R)_{12}(U_R)_{23}(U_R)_{13}^*(U_R)_{33}^*    )<3.6\times  10^{-6}  \nonumber  \\
& \rightarrow& -f_2 f_3 s_{12}(N_R)s_{23}(N_R)s_{13}(N_R)c^3_{13}(N_R)c_{23}(N_R)sin\delta \nonumber \\
&& <3.6\times  10^{-6} .
\end{eqnarray}
As mentioned above, we assume  $sin\theta_{12}(N_R)\sim sin\theta_{23}(N_R)\sim sin\theta_{13}(N_R)=sin\alpha =-\sqrt{3/2}sin\theta_{13}   $, 
where the last equality comes from Eq. (16) and  $ cos\alpha \sim1 $. Then Eq. (28) will become
\begin{eqnarray}
 f_2 f_3sin^3\theta_{13} sin\delta <1.96\times  10^{-6}.
\end{eqnarray}
Taking  $sin \theta_{13} \approx 0.1503  $, Eq. (29) will become
\begin{eqnarray}
f_2 f_3 sin\delta <5.77\times  10^{-4}.
\end{eqnarray}
Choosing  $f_2 f_3 \sim 10^{-2}$, we will obtain $ sin\delta <0.0577 $  which is in the range of NOvA experiment \cite{nova}.

We  added small breaking terms $M_{ij}N_{R_i}N_{R_j}  $  in model based on $A_4$ symmetry to  generate non-diagonal right-handed neutrino mixing, which gave a small correction to TBM mixing via seesaw mechanism with nonzero reactor angle and CP violating phase. We estimated the CP violating phase by investigating the process of leptogenesis due to the decay of right-handed neutrino. And the predicted CP violating phase was in the range of NOvA experiment \cite{nova}.

\begin{widetext}
\section*{Appendix}
\subsection{ $S_4$ symmetry   }

 $S_4$  group covers all permutations among four objects  $(x_1, x_2, x_3, x_4     )    $, i.e.  $(x_1, x_2, x_3, x_4     )\rightarrow  (x_i, x_j, x_k, x_l     )   $ and includes 24 elements as in literature \cite{nonabelian}
\begin{eqnarray}
&&a_1:(x_1, x_2, x_3, x_4     ),\quad a_2:(x_2, x_1, x_4, x_3     ),\quad a_3:(x_3, x_4, x_1, x_2     ),\quad a_4:(x_4, x_3, x_2, x_1     ), \nonumber   \\
&&b_1:(x_1, x_4, x_2, x_3     ),\quad b_2:(x_4, x_1, x_3, x_2     ),\quad b_3:(x_2, x_3, x_1, x_4     ),\quad b_4:(x_3, x_2, x_4, x_1     ), \nonumber   \\
&&c_1:(x_1, x_3, x_4, x_2     ),\quad c_2:(x_3, x_1, x_2, x_4     ),\quad c_3:(x_4, x_2, x_1, x_3     ),\quad c_4:(x_2, x_4, x_3, x_1     ), \nonumber   \\
&&d_1:(x_1, x_2, x_4, x_3     ),\quad d_2:(x_2, x_1, x_3, x_4     ),\quad d_3:(x_4, x_3, x_1, x_2     ),\quad d_4:(x_3, x_4, x_2, x_1     ), \nonumber   \\
&&e_1:(x_1, x_3, x_2, x_4     ),\quad e_2:(x_3, x_1, x_4, x_2     ),\quad e_3:(x_2, x_4, x_1, x_3     ),\quad e_4:(x_4, x_2, x_3, x_1     ), \nonumber   \\
&&f_1:(x_1, x_4, x_3, x_2     ),\quad f_2:(x_4, x_1, x_2, x_3    ),\quad f_3:(x_3, x_2, x_1, x_4     ),\quad f_4:(x_2, x_3, x_4, x_1    )
\end{eqnarray}
which are written in cycle representation as
\begin{eqnarray}
&&a_1:e,\quad a_2:(12)(34),\quad a_3:(13)(24),\quad a_4:(14)(32), \nonumber   \\
&&b_1:(234),\quad b_2:(124),\quad b_3:(132),\quad b_4:(314), \nonumber   \\
&&c_1:(243),\quad c_2:(231),\quad c_3:(134),\quad c_4:(142), \nonumber   \\
&&d_1:(34),\quad d_2:(12),\quad d_3:(1324),\quad d_4:(2314), \nonumber   \\
&&e_1:(23),\quad e_2:(2431),\quad e_3:(1342),\quad e_4:(14), \nonumber   \\
&&f_1:(24),\quad f_2:(1234),\quad f_3:(13),\quad f_4:(1432)
\end{eqnarray}
$S_4$ is cube symmetry group and the elements can also be written as
\begin{eqnarray}
&&a_1=\left(\begin{array}{ccc}
 1 & 0 & 0 \\
0 &1 &0\\
0&0  & 1
 \end{array} \right),\quad a_2=\left(\begin{array}{ccc}
 1 & 0 & 0 \\
0 &-1 &0\\
0&0  & -1
 \end{array} \right),\quad a_3=\left(\begin{array}{ccc}
 -1 & 0 & 0 \\
0 &1 &0\\
0&0  & -1
 \end{array} \right),\quad a_4=\left(\begin{array}{ccc}
 -1 & 0 & 0 \\
0 &-1 &0\\
0&0  & 1
 \end{array} \right), \nonumber   \\
&&b_1=\left(\begin{array}{ccc}
 0 & 0 & 1 \\
1 &0 &0\\
0&1  & 0
 \end{array} \right),\quad b_2=\left(\begin{array}{ccc}
  0 & 0 & 1 \\
-1 &0 &0\\
0&-1  & 0
 \end{array} \right),\quad b_3=\left(\begin{array}{ccc}
 0 & 0 & -1 \\
1 &0 &0\\
0&-1  & 0
 \end{array} \right),\quad b_4=\left(\begin{array}{ccc}
  0 & 0 & -1 \\
-1 &0 &0\\
0&1  & 0
 \end{array} \right), \nonumber   \\
&&c_1=\left(\begin{array}{ccc}
 0 & 1 & 0 \\
0 &0 &1\\
1&0  & 0
 \end{array} \right),\quad c_2=\left(\begin{array}{ccc}
  0 & 1 & 0 \\
0 &0 &-1\\
-1&0  & 0
 \end{array} \right),\quad c_3=\left(\begin{array}{ccc}
  0 & -1 & 0 \\
0 &0 &1\\
-1&0  & 0
 \end{array} \right),\quad c_4=\left(\begin{array}{ccc}
  0 & -1 & 0 \\
0 &0 &-1\\
1&0  & 0
 \end{array} \right), \nonumber   \\
&&d_1=\left(\begin{array}{ccc}
 1 & 0 & 0 \\
0 &0 &1\\
0&1  & 0
 \end{array} \right),\quad d_2=\left(\begin{array}{ccc}
 1 & 0 & 0 \\
0 &0 &-1\\
0&-1  & 0
 \end{array} \right),\quad d_3=\left(\begin{array}{ccc}
 -1 & 0 & 0 \\
0 &0 &1\\
0&-1  & 0
 \end{array} \right),\quad d_4=\left(\begin{array}{ccc}
 -1 & 0 & 0 \\
0 &0 &-1\\
0&1  & 0
 \end{array} \right), \nonumber   \\
&&e_1=\left(\begin{array}{ccc}
 0 & 1 & 0 \\
1 &0 &0\\
0&0  & 1
 \end{array} \right),\quad e_2=\left(\begin{array}{ccc}
0 & 1 & 0 \\
-1 &0 &0\\
0&0  & -1
 \end{array} \right),\quad e_3=\left(\begin{array}{ccc}
 0 & -1 & 0 \\
1 &0 &0\\
0&0  & -1
 \end{array} \right),\quad e_4=\left(\begin{array}{ccc}
0 & -1 & 0 \\
-1 &0 &0\\
0&0  & 1
 \end{array} \right), \nonumber   \\
&&f_1=\left(\begin{array}{ccc}
 0 & 0 & 1 \\
0 &1 &0\\
1&0  & 0
 \end{array} \right),\quad f_2=\left(\begin{array}{ccc}
 0 & 0 & 1 \\
0 &-1 &0\\
-1&0  & 0
 \end{array} \right),\quad f_3=\left(\begin{array}{ccc}
 0 & 0 & -1 \\
0 &1 &0\\
-1&0  & 0
 \end{array} \right),\quad f_4=\left(\begin{array}{ccc}
 0 & 0 & -1 \\
0 &-1 &0\\
1&0  & 0
 \end{array} \right)
\end{eqnarray}

Generally elements are classified by symmetry operations as follows:  $T$ represents element generated by rotating around coordinate axes which including 3 axes  $x,y,z$: rotating $\pi$  for 2-order element and $\pm \frac{1}{2}\pi$ for 4-order element; $S$  represents 2-order element generated by rotating around axes connecting the midpoints in two opposite edges which include 6 axes.  $R$  represents  3-order element generated by rotating around body diagonals which include 4 axes. The elements are classified by axes as follows
\begin{eqnarray}
&&\textrm{T\ axis}: \{ a_2, a_3, a_4, d_3, d_4, e_2, e_3, f_2, f_4      \} \nonumber   \\
&&\textrm{R\ axis}: \{ b_1, b_2, b_3, b_4, c_1, c_2, c_3, c_4     \} \nonumber   \\
&&\textrm{S\ axis}: \{ d_1, d_2, e_1, e_4, f_1, f_3     \} 
\end{eqnarray}
and  by the order of element as
\begin{eqnarray}
&&h=1: \{ a_1    \} \nonumber   \\
&&h=2: \{ a_2, a_3, a_4, d_1, d_2, e_1, e_4, f_1, f_3     \} \nonumber   \\
&&h=3: \{ b_1, b_2, b_3, b_4, c_1, c_2, c_3, c_4     \} \nonumber   \\
&&h=4: \{ d_3, d_4, e_2, e_3, f_2, f_4     \} 
\end{eqnarray}
Elements  are classified into conjugate classes as
\begin{eqnarray}
&&C_1(h=1): \{ a_1    \} \nonumber   \\
&&C_3(h=2): \{ a_2, a_3, a_4     \} \nonumber   \\
&&C_6(h=2): \{  d_1, d_2, e_1, e_4, f_1, f_3     \} \nonumber   \\
&&C_8(h=3): \{ b_1, b_2, b_3, b_4, c_1, c_2, c_3, c_4     \} \nonumber   \\
&&C_6'(h=4): \{ d_3, d_4, e_2, e_3, f_2, f_4     \} 
\end{eqnarray}

Permutation group or symmetric group $S_n$ includes two generators which could be chosen one cycle of n length, e.g. $(123\cdots n) $  and one cycle of its neighboring objects e.g.  $(12)$. It could be easily verified by proving $(j\quad j+1)=(123\cdots n)(j-1\quad j)(123\cdots n)^{-1}$. For example, from Eq. (32), we could choose $d_1$ and corresponding n-length cycle $f_2$  as generators of  $S_4$.

\subsection{$A_4$ symmetry    }

 $A_4$  group covers all even permutations of four objects, which is subgroup of $S_4$  and includes 12 elements as follows
\begin{eqnarray}
&&a_1=\left(\begin{array}{ccc}
 1 & 0 & 0 \\
0 &1 &0\\
0&0  & 1
 \end{array} \right),\quad a_2=\left(\begin{array}{ccc}
 1 & 0 & 0 \\
0 &-1 &0\\
0&0  & -1
 \end{array} \right),\quad a_3=\left(\begin{array}{ccc}
 -1 & 0 & 0 \\
0 &1 &0\\
0&0  & -1
 \end{array} \right),\quad a_4=\left(\begin{array}{ccc}
 -1 & 0 & 0 \\
0 &-1 &0\\
0&0  & 1
 \end{array} \right), \nonumber   \\
&&b_1=\left(\begin{array}{ccc}
 0 & 0 & 1 \\
1 &0 &0\\
0&1  & 0
 \end{array} \right),\quad b_2=\left(\begin{array}{ccc}
  0 & 0 & 1 \\
-1 &0 &0\\
0&-1  & 0
 \end{array} \right),\quad b_3=\left(\begin{array}{ccc}
 0 & 0 & -1 \\
1 &0 &0\\
0&-1  & 0
 \end{array} \right),\quad b_4=\left(\begin{array}{ccc}
  0 & 0 & -1 \\
-1 &0 &0\\
0&1  & 0
 \end{array} \right), \nonumber   \\
&&c_1=\left(\begin{array}{ccc}
 0 & 1 & 0 \\
0 &0 &1\\
1&0  & 0
 \end{array} \right),\quad c_2=\left(\begin{array}{ccc}
  0 & 1 & 0 \\
0 &0 &-1\\
-1&0  & 0
 \end{array} \right),\quad c_3=\left(\begin{array}{ccc}
  0 & -1 & 0 \\
0 &0 &1\\
-1&0  & 0
 \end{array} \right),\quad c_4=\left(\begin{array}{ccc}
  0 & -1 & 0 \\
0 &0 &-1\\
1&0  & 0
 \end{array} \right)
\end{eqnarray}
$A_4$  group is the symmetry group of regular tetrahedron which generators generally are called   $S,T$. $T$ represents 3-order element generated by  rotating around axes passing through vertices which include 4 axes, which can be classified as: element by rotating $ \frac{2}{3}\pi  $  clockwise and element by rotating $ \frac{2}{3}\pi  $  anticlockwise. $S$  represents 2-order element generated by rotating around axes connecting the midpoints in two opposite edges which include 3 axes.

All 12 elements can be classified by axes as follows
\begin{eqnarray}
&&\textrm{S\ axis}: \{ a_2, a_3, a_4   \} \nonumber   \\
&&\textrm{T\ axis}: \{ b_1, b_2, b_3, b_4, c_1, c_2, c_3, c_4     \}
\end{eqnarray}
The  generators could be chosen  $ S=a_2, T= b_1  $  which satisfy $S^2 = T^3 =1   $.

More can be referred to literature \cite{nonabelian}.

\end{widetext}

\end{document}